# LA ORIENTACIÓN DE LAS IGLESIAS COLONIALES DE FUERTEVENTURA

# THE ORIENTATION OF COLONIAL CHURCHES OF FUERTEVENTURA

M.F. Muratore [a], A. Gangui [b],∗

a Universidad Nacional de Luján, Departamento de Ciencias Básicas, Buenos Aires, Argentina. CONICET – Universidad de Buenos Aires, Instituto de Astronomía y Física del Espacio (IAFE), Argentina
b Universidad de Buenos Aires, Facultad de Ciencias Exactas y Naturales, Argentina. CONICET – Universidad de Buenos Aires, Instituto de Astronomía y Física del Espacio (IAFE), Argentina

Empleamos las herramientas usuales de la Arqueoastronomía para abordar el estudio de orientaciones, eventualmente astronómicas, de un grupo de iglesias cristianas coloniales. Presentamos resultados preliminares del análisis de la orientación espacial precisa de medio centenar de ermitas e iglesias de la isla canaria de Fuerteventura (España), la mayoría de ellas construidas desde el período de la conquista normanda en el siglo XV hasta el siglo XIX. A pesar de que algunas pequeñas ermitas pertenecientes al poder señorial de la isla y otras iglesias modernas no presentan un patrón de orientaciones bien definido, la gran mayoría de las construcciones religiosas de la isla (unas 35 de las 48 analizadas) posee sus ejes orientados dentro del rango solar, entre los acimuts extremos del movimiento anual del Sol al cruzar el horizonte local. A diferencia de lo que fue hallado en otras islas del archipiélago, estos resultados sugieren que la arquitectura religiosa de Fuerteventura sigue fielmente las prescripciones contenidas en los textos de los escritores cristianos tempranos.

*Palabras clave: orientación de iglesias, templos cristianos, Astronomía.*

We use standard tools of Archaeoastronomy to approach the study of orientations, possibly astronomical, of a group of colonial Christian churches. We present preliminary results of the analysis of the precise spatial orientation of nearly fifty chapels and churches of the Canary Island of Fuerteventura (Spain), most of them built from the period of the Norman conquest in the fifteenth century to the nineteenth century. Although some small chapels belonging to the manorial power of the island and other modern churches do not have a well-defined pattern of orientations, the vast majority of the religious constructions of the island (about 35 of the 48 analyzed) have their axis oriented within the solar range, between the extreme azimuths of the annual movement of the Sun when crossing the local horizon. Unlike what was found in other islands of the archipelago, these results suggest that the religious architecture of Fuerteventura faithfully follows the prescriptions contained in the texts of early Christian writers.

*Keywords: church orientation, Christian religion, Astronomy.*

## I. INTRODUCCIÓN

Desde sus inicios, la Arqueoastronomía ha concentrado sus principales esfuerzos en el estudio de construcciones históricas -como los megalitos europeos, las pirámides egipcias o los templos mesoamericanos- y en indagar las posibles influencias de los cuerpos celestes en sus estructuras y diseños. El estudio de orientaciones de las iglesias medievales fue también, desde siempre, otro de los objetivos más frecuentados. En este último tema, una multitud de trabajos recientes (por ejemplo, el de González-García y Belmonte[1]) muestra que las prescripciones para la orientación de los ábsides hacia el oriente se siguieron de forma muy sistemática en toda Europa, al menos durante la Edad Media (Fig. 1).

A partir de textos antiguos sabemos que la orientación espacial de las iglesias cristianas históricas es una de las características destacadas de su arquitectura,[2] con una visible tendencia a orientar los altares de los templos en el rango solar. Así, el eje principal del templo, desde la puerta de entrada hacia el altar, debe alinearse con los puntos en el horizonte por donde sale el Sol en diferentes días del año. Entre estos días, hay una marcada preferencia por los que corresponden a los equinoccios astronómicos, cuando los ejes apuntan hacia el este geográfico.[3]

Aunque los investigadores se han dedicado preferentemente a analizar iglesias particulares de las islas británicas y de Europa continental, concentrándose en sus orientaciones y eventuales eventos de iluminación, poco a poco se han ido desarrollando estudios sobre la orientación de los templos en períodos posteriores a la Edad Media y en regiones alejadas del centro europeo. Es en este contexto en el que se inscribe nuestro estudio presente, pues como veremos una gran mayoría de las iglesias y ermitas de la isla canaria de Fuerteventura comenzó a erigirse décadas después de la conquista y colonización de la isla por los expedicionarios normandos que contaban con el beneplácito de la corona de Castilla en los inicios del siglo XV.[4]

---

∗ *flormuratore@gmail.com*

En este trabajo ofrecemos un primer análisis de nuestros datos -recabados recientemente en un trabajo de campo de medición de orientaciones, posiblemente astronómicas- de iglesias coloniales históricas ubicadas en Fuerteventura. En continuidad con trabajos anteriores ya desarrollados en islas vecinas, nuestra intención es indagar si en este territorio acotado, y ubicado lejos del gobierno central, se respetaron -o no- los textos de los escritores y apologetas cristianos tempranos en lo que respecta a la orientación *ad orientem* de la arquitectura religiosa.[5]

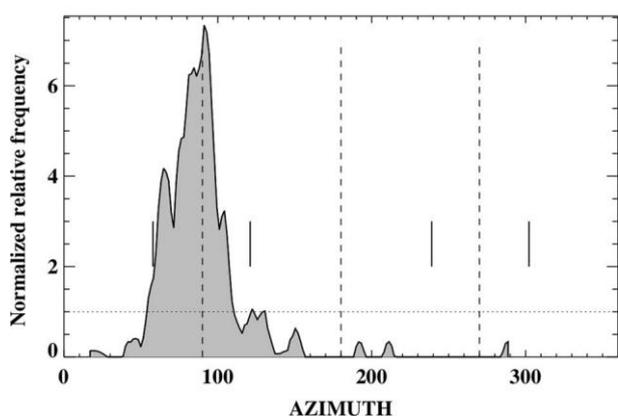

*Figura 1. Histograma de acimut de orientación de una muestra de 167 iglesias prerrománicas de España y Portugal construidas antes del año 1086 de nuestra era.[1] El gráfico muestra una clara tendencia de estas iglesias medievales a orientar sus ejes hacia el este geográfico (acimut de 90º), correspondiente al equinoccio astronómico.*

## II. LAS ERMITAS E IGLESIAS DE FUERTEVENTURA

Las primeras construcciones religiosas de Fuerteventura fueron pequeñas ermitas dotadas de un único recinto y de factura sencilla. Un ejemplo es la ermita de San Isidro Labrador, edificada en el año 1714 en el pueblo de Triquivijate, dentro del municipio de Antigua (Fig. 2). Esta, como tantas otras, está dotada de una planta rectangular, posee una fachada de frente plano y hoy está cubierta por un tejado a tres aguas y teja al exterior. Muchas de estas ermitas cuentan con más de una puerta de acceso, como San Isidro, cuya entrada lateral está abierta hacia el cuadrante sur, y tanto esta última como la portada principal poseen sus vanos realizados en cantería clara que vienen rematados en arcos de medio punto.

A algunas de estas ermitas, con el correr de los años, se les fueron agregando capillas en la cabecera, sacristías a sus lados y otros elementos de uso práctico. San Isidro, por ejemplo, cuenta en el lado derecho superior de su fachada con una pequeña espadaña dotada de un único hueco, que hace las veces de modesto campanario, realizada en cantería clara. Y en su lateral izquierdo una pequeña sacristía se adosa al muro del evangelio, donde se comunica con la nave principal (Fig. 2).

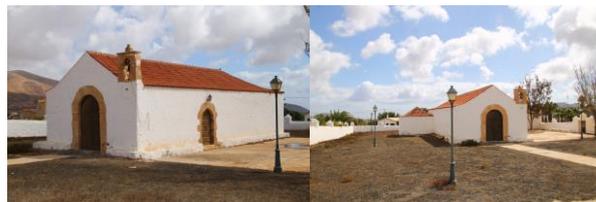

*Figura 2. La ermita de San Isidro Labrador, en Triquivijate, es una de las construcciones típicas de los templos cristianos de la isla. Una pequeña sacristía se conecta con la nave principal de la ermita, cuya fachada muestra una modesta espadaña que oficia de campanario.*

El pequeño templo de San Isidro Labrador, al igual que tantos que se construyeron en Fuerteventura, fue sufragado por el pueblo. Y como este, muchos no estuvieron sujetos a planes de ejecución estrictos. Por ello, tanto su planta como su estructura se levantaron -y con los años fueron creciendo- de acuerdo con las necesidades del momento. Con el tiempo, algunas ermitas se ampliaron lo suficiente como para adquirir un cierto carácter monumental, con portadas dotadas de grandes arcos de medio punto, espadañas de uno o varios vanos con varias campanas y techos a dos o cuatro aguas, la mayoría de las veces con tejas.

En aquellas ocasiones en que las iglesias no han sido aún oprimidas por la ciudad que crece a su alrededor, los recintos permanecen rodeados por un amplio muro exterior almenado o barbacana, como es el caso no solo de San Isidro (Fig. 3), sino también el de San Antonio de Padua en Lajares, el de Nuestra Señora de Guadalupe en Agua de Bueyes, y el de varios otros templos de la isla. En el caso particular de la ermita de Agua de Bueyes, sabemos que las puertas en su barbacana solo fueron instaladas en 1792 con el fin preciso de evitar la entrada de ganado.[6]

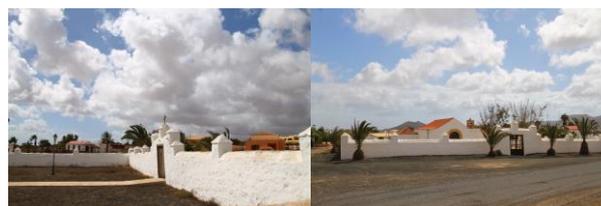

*Figura 3. La extensa barbacana que rodea a la ermita de San Isidro cuenta con dos entradas con puertas de madera que enfrentan a los accesos al interior del templo.*

Dado el elevado número de estas construcciones históricas, cercano a la media centena y por lo tanto estadísticamente significativo, tenida en cuenta la pequeña extensión del territorio, de algo menos de 1700 km² (por hacer una comparación, menos de la mitad de la superficie de la isla San Pedro, en las Georgias del Sur), decidimos trabajar en Fuerteventura como un nuevo caso de estudio en donde testear la orientación de las iglesias canarias en los siglos inmediatamente posteriores a la conquista europea. Este estudio es la continuación natural de un proyecto a gran escala que en los últimos años se ha venido concentrando en la medición sistemática de iglesias y otros sitios de culto, dentro del marco de la arqueología del paisaje, tanto en la Península Ibérica como en otras regiones de Europa y en particular en las Canarias.[7,3,8,9] Parte de nuestra motivación es analizar si en la orientación de las ermitas e iglesias en este territorio ha habido influencias de elementos propios de la población aborigen prehispánica (los maxos o majoreros, de origen bereber), que en principio habría tenido unos patrones de culto diferentes de aquellos de los colonizadores recién llegados.[10]

En la Tabla 1 presentamos nuestros datos, obtenidos en una campaña de trabajo de campo donde se cubrió la totalidad de la superficie de la isla. En esta incluimos la información estándar de cada iglesia o ermita, es decir, su identificación y sus coordenadas geográficas junto con su orientación, dada por el acimut de los ejes de las construcciones medido en cada sitio, y luego corregido de acuerdo con la declinación magnética local. Por último, se incluye la altura angular del horizonte tomada a lo largo del eje de cada edificio en dirección al altar. Para aquellas iglesias cuyos horizontes en la cabecera estaban bloqueados, por ejemplo, por edificaciones modernas (señaladas con B en la última columna de la Tabla), planeamos realizar una futura reconstrucción del horizonte usando el modelo digital del terreno disponible en heywhatsthat.com.

Nuestros datos se obtuvieron con brújulas de alta precisión. Los valores de la declinación magnética para distintos sitios de la isla oscilan entre 4º14' y 4º21' oeste. La precisión de nuestras medidas de acimut magnéticos es de aproximadamente 0,5º, por lo que la diferencia en declinación magnética a lo largo de la isla entra adecuadamente dentro de nuestro error. Como una corroboración adicional, se verificó la totalidad de las orientaciones medidas con imágenes fotosatelitales.

En la Fig. 5 mostramos el diagrama de orientación para las iglesias y ermitas analizadas. Los valores de los acimuts consignados son los medidos, e incluyen la corrección por declinación magnética en cada sitio. Las líneas diagonales del gráfico señalan los acimuts correspondientes -en el cuadrante oriental- a los valores extremos para el Sol (acimuts de 62,7º y 116,6º -líneas continuas-, equivalente a los solsticios de verano e invierno boreales, respectivamente) y para la Luna (acimuts: 56,6º y 123,6º -líneas rayadas-, equivalente a la posición de los lunasticios mayores).

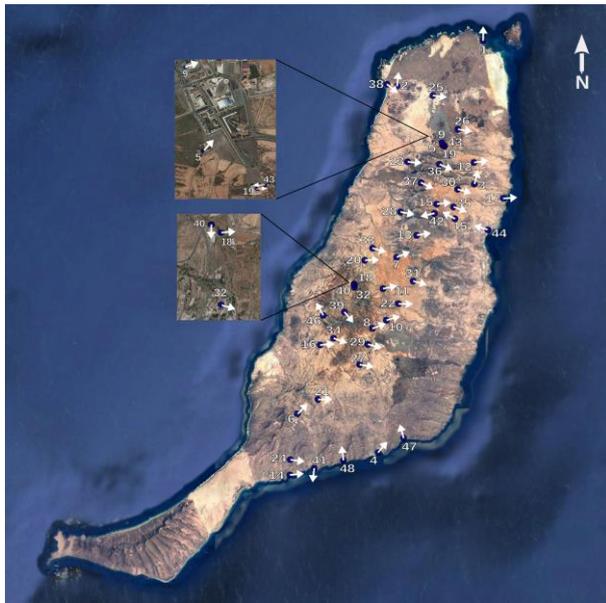

*Figura 4. Mapa con la ubicación geográfica de la totalidad de las iglesias medidas (señaladas con círculos oscuros), junto con la orientación del eje de las construcciones en dirección al altar (flechas, apuntando de acuerdo con los acimuts consignados en la Tabla 1). En los alrededores de la ciudad de La Oliva, hacia el norte de la isla, hay cuatro construcciones geográficamente muy próximas (la iglesia de La Candelaria y tres ermitas, todas con orientaciones diferentes), lo que explica la presencia allí de círculos solapados (ampliamos la región en la figura insertada, incluyendo las cuatro flechas). Algo similar ocurre en la Villa de Betancuria, en la región central, donde una iglesia, un convento y una ermita se ubican a pocas centenas de metros y poseen orientaciones muy diferentes. Imagen de los autores sobre un mapa cortesía de Google Earth.*

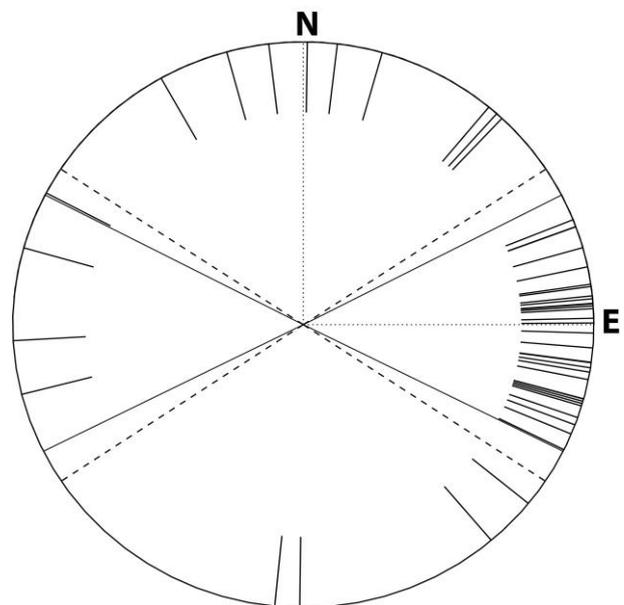

*Figura 5. Diagrama de orientación para las iglesias y ermitas de Fuerteventura, obtenido a partir de los datos de la Tabla 1.*

A pesar de que algunas ermitas pequeñas y/o modernas se hallan orientadas en los cuadrantes norte y sur, y por lo tanto se ubican fuera del rango solar, la gran mayoría de las construcciones religiosas de la isla (unas 35) siguen la orientación canónica acomodando sus ejes entre los acimuts extremos del movimiento anual del Sol al cruzar el horizonte local. Este resultado, si bien es coherente con el patrón de orientaciones de grupos de iglesias de épocas anteriores típicas de los lugares de origen de los colonizadores (Ref. 5; ver Fig. 1), difiere notablemente de los resultados obtenidos en otras islas del archipiélago canario y requiere una explicación.

Tabla 1: Orientaciones de las iglesias de la isla de Fuerteventura, ordenadas por acimut creciente. Para cada construcción, la tabla muestra la ubicación, la identificación, la latitud y longitud geográficas (L y l), el acimut astronómico (a) y la altura angular del horizonte medida (h), sin corrección por refracción atmosférica, tomados a lo largo del eje del edificio en dirección al altar (aproximados al 0.5° de error), con valores expresados en grados decimales. En la última columna, B señala que el horizonte estaba bloqueado y por ello tomamos h = 0°. Los números de la primera columna corresponden con aquellos anotados en la Fig. 4 del mapa con sus ubicaciones geográficas.

| Ubicación | Nombre | L (°, N) | l (°, O) | a (°) | h (°) |
|---|---|---|---|---|---|
| (1) Corralejo | Ntra. Sra. del Carmen | 28.742530 | 13.867266 | 1.0 | 0 |
| (2) El Roque | San Martín de Porres | 28.684159 | 13.994220 | 6.5 | B 0 |
| (3) Guisguey | San Pedro Apóstol | 28.557882 | 13.880684 | 16.0 | +8 |
| (4) Gran Tarajal | Ntra. Sra. de la Candelaria | 28.212097 | 14.021827 | 39.5 | B 0 |
| (5) La Oliva | Ntra. Sra. De Puerto Rico (Capellanía) | 28.607969 | 13.927410 | 42.0 | +2.5 |
| (6) Cardón | Ermita de Cardón | 28.262220 | 14.140538 | 43.0 | +7.5 |
| (7) La Ampuyenta | San Pedro de Alcántara | 28.463512 | 13.995226 | 68.0 | +6.5 |
| (8) Agua de Bueyes | Ntra. Sra. de Guadalupe | 28.372139 | 14.030763 | 69.5 | +2.5 |
| (9) La Oliva | Ntra. Sra. de la Candelaria | 28.611131 | 13.928146 | 70.0 | +3 |
| (10) Valle de Ortega | San Roque | 28.382575 | 14.010252 | 74.0 | +1 |
| (11) Antigua | Ntra. Sra. de Antigua | 28.423258 | 14.015106 | 78.0 | B 0 |
| (12) La Caldereta | Ntra. Sra. de los Dolores | 28.585625 | 13.881228 | 82.0 | 0 |
| (13) Casillas del Ángel | Santa Ana | 28.491843 | 13.965658 | 82.0 | +3.5 |
| (14) La Lajita | N. Sra. de la Inmaculada Concepción | 28.183281 | 14.151836 | 84.0 | +5 |
| (15) Tetir | Sto. Domingo de Guzmán | 28.531875 | 13.936238 | 85.0 | -0.5 |
| (16) Pájara | Ntra. Sra. de Regla | 28.350808 | 14.107416 | 85.5 | +4 |
| (17) Puerto Lajas | Virgen del Pino | 28.539775 | 13.837955 | 86.0 | 0 |
| (18) Betancuria | Iglesia Convento de San Buenaventura | 28.428687 | 14.057293 | 86.0 | +13 |
| (19) La Oliva | Ermita interna, Casa de los coroneles | 28.606554 | 13.925092 | 87.0 | +1 |
| (20) Valle Santa Inés | Santa Inés | 28.459603 | 14.042009 | 88.5 | +1 |
| (21) Tesejerague | San José | 28.280637 | 14.110490 | 89.5 | +10.5 |
| (22) Las Pocetas | San Francisco Javier | 28.403953 | 13.992415 | 91.5 | +2.5 |
| (23) Tindaya | Ntra. Sra. de la Caridad | 28.586484 | 13.978857 | 94.5 | +7 |
| (24) Tarajal de Sancho | Ntra. Sra. de Fátima | 28.203085 | 14.151140 | 97.5 | +7.5 |
| (25) Lajares | San Antonio de Padua | 28.672077 | 13.941469 | 98.0 | +1 |
| (26) Villaverde | San Vicente Ferrer | 28.628185 | 13.904735 | 99.0 | +8.5 |
| (27) Tuineje | San Miguel Arcángel | 28.325541 | 14.049361 | 99.5 | B 0 |
| (28) Tefía | San Agustín | 28.522016 | 13.989525 | 101.0 | +8 |
| (29) Tiscamanita | San Marcos Evangelista | 28.351585 | 14.037200 | 105.0 | 0 |
| (30) El Time | Ntra. Sra. de la Merced | 28.550914 | 13.905613 | 105.5 | -0.5 |
| (31) Triquivijate | San Isidro Labrador | 28.433100 | 13.970822 | 105.5 | 0 |
| (32) Betancuria | Santa María | 28.425022 | 14.057285 | 106.0 | +8 |
| (33) Llanos de la Concepción | Ntra. Sra. de la Inmaculada Concepción | 28.475294 | 14.030411 | 106.5 | +2 |
| (34) Toto | San Antonio de Padua | 28.359578 | 14.087736 | 109.0 | +2.5 |
| (35) Los Estancos | Santa Rita | 28.528229 | 13.911651 | 111.0 | 0 |
| (36) Vallebrón | San Juan Bautista | 28.583180 | 13.932058 | 113.0 | +4 |
| (37) La Matilla | Ntra. Sra. del Socorro | 28.559390 | 13.959246 | 116.0 | +1.5 |
| (38) El Cotillo | Ntra. Sra. del Buen Viaje | 28.686680 | 14.009098 | 132.5 | +2 |
| (39) Vega de Río Palmas | Ntra. Sra. de la Peña | 28.393220 | 14.071690 | 139.5 | +10.5 |
| (40) Betancuria | San Diego de Alcalá | 28.429098 | 14.057783 | 180.5 | +2.5 |
| (41) Tarajalejo | Ermita de Tarajalejo | 28.191968 | 14.115993 | 185.5 | 0 |
| (42) Tetir | San Andrés | 28.520150 | 13.939539 | 255.5 | +6 |
| (43) La Oliva | Ermita exterior, Casa de los coroneles | 28.606662 | 13.924726 | 267.0 | 0 |
| (44) Pto. del Rosario | Ntra. Sra. del Rosario | 28.498975 | 13.860517 | 286.0 | +3 |
| (45) La Asomada | Ntra. Sra. de Fátima | 28.513017 | 13.909287 | 298.0 | +1.5 |
| (46) Bco de las Peñitas | Ermita del Malpaso | 28.388833 | 14.102847 | 330.5 | +38.5 |
| (47) Las Playitas | San Pedro Apóstol | 28.230954 | 13.984837 | 344.5 | +2 |
| (48) Giniginamar | Ntra. Sra. del Carmen | 28.201998 | 14.073144 | 353.0 | +3.5 |

## III. ORIENTACIÓN CANÓNICA PERO SINGULAR EN CANARIAS

De las 48 orientaciones medidas en las ermitas e iglesias, 9 se dirigen hacia el cuadrante norte, 3 hacia el meridional, 4 hacia el occidental y la gran mayoría (32) apuntan hacia el cuadrante oriental. De estos dos últimos grupos, hay 35 construcciones cuyas orientaciones se ubican en el rango solar, ya sea a levante (31) o a poniente (4).

Nuestra muestra es representativa de toda la isla de Fuerteventura, y estos datos nos permiten inferir un cierto patrón de orientación. Del diagrama de la Fig. 5 se distingue una orientación dominante en el rango solar hacia levante. Como señalamos, este es un patrón esperable de acuerdo con los textos de los escritores y apologetas cristianos tempranos. Sin embargo, esto difiere notoriamente del patrón hallado previamente en otras dos islas ya estudiadas en Canarias: difiere de los resultados de Lanzarote, donde una notable proporción de las iglesias se orientaba aproximadamente hacia el norte-noreste (con "entrada" a sotavento) para evitar los vientos alisios dominantes del lugar.[3] Y también difiere de lo hallado en La Gomera, donde se halló un patrón de orientación aproximadamente similar al de Lanzarote, pero debido a una razón diferente. Se ha visto que en La Gomera la acumulación de orientaciones en el cuadrante noreste se debe a las características orográficas de la isla, donde varios grupos de templos se orientan siguiendo la dirección de un par de valles profundos entre aquellos que marcan su "abrupta geografía"[11] y que se dirigen aproximadamente del sudoeste hacia el noreste.[9]

Por el momento, entonces, nuestro estudio preliminar ubica a la isla de Fuerteventura en una posición "singular" entre los demás miembros del archipiélago canario ya analizados, ya que los factores "prácticos" y climáticos preponderantes en la vecina Lanzarote, u orográficos, determinantes en La Gomera, no parecen estar presentes en esta isla, la más cercana al África continental.

Dentro del grupo de iglesias históricas que poseen una orientación canónica (en el rango solar) se hallan las quizá más representativas de la isla, como ser la iglesia Santa María de Betancuria (número 32 en la Tabla), parroquia matriz de la Villa homónima y de todo Fuerteventura, cuya construcción fue instruida por el mismo expedicionario normando Jean de Béthencourt, conquistador de la isla, alrededor de 1410. Su eje principal se orienta con un acimut de 106,0°, lo que la ubica a un poco más de una decena de grados hacia el sur del este geográfico. En la misma Villa capital, aunque de fecha un poco posterior, se halla la iglesia del Convento Franciscano de San Buenaventura (núm 18), del que solo resta la sólida estructura y los muros de una reconstrucción del siglo XVII, posterior a los ataques berberiscos, ya que su tejado no sobrevivió. Esta estructura de la iglesia conventual también está orientada hacia levante con un acimut de 86,0° y a escasos grados al norte del este.

Otras iglesias históricas con orientaciones muy próximas al este son Santa Ana, en Casillas del Ángel (núm 13), cuyo eje tiene un acimut de 82,0° y, en la localidad de Lajares, la ermita de San Antonio de Padua (núm 25), con un acimut de 98,0°. Ambas construcciones datan de la última parte del siglo XVIII[12] y distan menos de 10 grados del este geográfico. Por último, la Tabla nos muestra que tanto la ermita de San José en Tesejerague (núm 21; primera mitad del s. XVIII) como la de San Francisco Javier en Las Pocetas (núm 22; del año 1771, cf. Ref. 6) tienen sus ejes orientados con acimuts compatibles (en aproximadamente dos grados) con la dirección equinoccial.

Por otra parte, hallamos también dos iglesias que podríamos llamar "solsticiales", una histórica y otra moderna. La ermita Nuestra Señora del Socorro, en La Matilla (núm 37),[6] data de 1716 y su eje se alinea con la dirección del orto solar durante el solsticio de invierno boreal (con un acimut aproximado de 116°). Mientras que el eje de la más reciente Nuestra Señora de Fátima, en La Asomada (núm 45), sigue la dirección del ocaso del Sol durante el solsticio de verano boreal (con un acimut de 298°, aproximadamente), aunque también podría estar orientada hacia la salida del Sol en el solsticio de invierno si, en lugar de considerar la dirección hacia el altar, tomamos la dirección contraria, hacia donde se abre la puerta del templo, como se ha visto que sucede en islas cercanas.[8] Sin embargo, por el momento estas son solo conjeturas, pues el número de construcciones con estas características es exiguo y no representativo del conjunto completo.

En lo que respecta a la decena de iglesias de la Tabla que se orientan en los cuadrantes norte y sur, y por ello lejos del rango solar, creemos que, para muchas de ellas, hay razones prácticas que podrían explicar sus orientaciones medidas, ya sea debido a sus orígenes como ermitas privadas de familias de cierta relevancia en la isla o bien por sus ubicaciones frente a litorales marinos. Quedan algunas, por supuesto, cuyas orientaciones por el momento no hemos podido comprender, pero que continuamos analizando con la ayuda de documentos originales, como "Libros de Fábrica" o "Libros de Cuentas" y demás fuentes secundarias.[13,14,6]

Entre las ermitas cuyas orientaciones se distribuyen dentro del cuadrante norte tenemos varias ubicadas en la región sur y costera de la isla. Tal es el caso de las ermitas Del Carmen, en Giniginamar (núm 48), y San Pedro Apóstol, en Las Playitas (núm 47). Ambas son de construcción moderna y abren sus puertas en dirección al mar, y por ello es natural que sus altares estén orientados hacia el norte. La ermita de Tarajalejo (núm 41), sin embargo, aunque también está ubicada en la costa sur y no lejos de Las Playitas, tiene su cabecera -y no su puerta- dirigida hacia la costa y es por ello que su acimut medido se ubica a pocos grados del sur. Otra iglesia moderna ubicada en la zona sur de la isla es Nuestra Señora de la Candelaria de Gran Tarajal (núm 4). Su eje se orienta hacia el noreste y fuera del rango solar, y muy probablemente sus constructores no repararon en la tradición arquitectónica religiosa al edificar este gran templo ya inserto en una gran ciudad.

En nuestra muestra hay varias ermitas que se apartan de las orientaciones canónicas, y creemos que esto en parte puede deberse a que fueron edificadas por particulares o por ser de factura muy reciente. Tal es el caso de la ermita de la Capellania en La Oliva (núm 5) que data de aproximadamente el año 1500 y que se piensa fue la antigua vivienda de un clérigo del lugar, y por lo tanto no deberíamos esperar que haya sido muy estricto en obedecer las orientaciones tradicionales. En lo que concierne a la iglesia de San Martin de Porres, en El Roque (núm 2), o a la ermita de Cardón (núm 6), orientadas en dirección del cuadrante norte, ambas son de construcción reciente, y por ello entran en esta categoría sin presentar demasiadas dudas. También la ermita de San Diego de Alcalá en la Villa de Betancuria (núm 40) puede entrar en la categoría de capilla privada. La tradición afirma que fue erigida en donde se hallaba una pequeña cueva, sitio frecuente de oración del santo a mediados del siglo XV. Su orientación es casi meridiana y muy alejada del rango solar.

En la Vega de Río Palmas y su continuación en el Barranco de las Peñitas tenemos dos templos que comparten una historia de varios siglos.[15] La ermita del Malpaso (núm 46) data del siglo XV y, como lo indica su nombre, es de muy difícil acceso. Allí permaneció durante un tiempo la imagen de la patrona de Fuerteventura, la Virgen de la Peña, que la tradición afirma que había sido traída desde Francia por el conquistador normando Jean de Béthencourt para Betancuria, pero que luego había sido escondida en las Peñitas para salvarla de los ataques berberiscos de fines del siglo XVI. Por la dificultad de acceder a esa ermita, en el siglo XVIII se construyó la iglesia de Nuestra Señora de la Peña (núm 39), en Vega de Río Palmas, para comodidad de los devotos (Fig. 6).

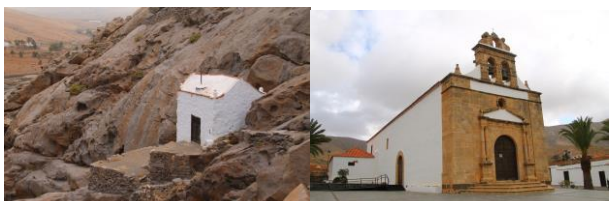

Figura 6. *La ermita del Malpaso (izquierda) situada en el barranco de las Peñitas, de difícil acceso, y la iglesia Nuestra Señora de la Peña (derecha), construida tiempo más tarde en Vega de Río Palmas, a pocos kilómetros de la primera. Por motivos diversos, sus ejes se orientan fuera del rango solar.*

Los ejes de ambas construcciones de esta vega se apartan mucho de las orientaciones canónicas. En el caso de la pequeña ermita, la inspección del paisaje circundante nos indica que el movimiento del Sol y el arco de sitios del horizonte por donde surge y se oculta durante el año fueron irrelevantes para sus artífices (y su alto valor de "h" en la Tabla así lo confirma), pues se halla ubicada en la ladera de un barranco profundo. Así, si bien es simple comprender que no hubo mucho margen de maniobras en la orientación de la ermita de las Peñitas, la orientación sudeste y fuera del rango solar de la iglesia posterior Nuestra Señora de la Peña, por el contrario, no fue adecuadamente documentada y es un tema que requiere una mayor exploración.

Quedan aún por explorar los templos de El Cotillo, Guisguey y Corralejo, todos ellos ubicados en la parte norte de la isla y con orientaciones fuera del rango solar. La ermita Nuestra Señora del Buen Viaje, en El Cotillo, se halla cercana a la costa occidental (núm 38) pero no hemos hallado referencia alguna (topográfica, por ejemplo) que justifique su orientación. Data de 1680 (Ref. 6, p. 322) pero su fachada fue restaurada en 1834, como figura en el frontis de la iglesia. Su acimut de 132,5° se aparta en más de 15 grados de la posición del Sol naciente del solsticio de invierno boreal y entre las fuentes documentales no hay datos que revelen los motivos de esta orientación. Por otra parte, tanto San Pedro Apóstol, en Guisguey, como Nuestra Señora del Carmen, en Corralejo, se orientan con acimuts próximos al norte geográfico (núm 3 y 1, respectivamente). San Pedro, en particular, está emplazada en un valle al pie de una sierra de mediana altura y la cabecera de la ermita mira hacia esa elevación del terreno. Sin embargo, aún no hemos podido hallar fuentes que detallen los planes de construcción originales. La iglesia de Corralejo, por su parte, es de construcción moderna, pero quedan registros que indican que la iglesia antigua, hoy desaparecida, habría estado emplazada en el mismo sitio y con aproximadamente la misma orientación que la moderna. Fotos antiguas sugieren que la iglesia original apuntaba su cabecera aproximadamente hacia el volcán Montaña Roja en la cercana isla de Lanzarote, pero aparte de esto no quedan indicios que justifiquen su orientación.

Dejamos para un estudio futuro el análisis orográfico del territorio y su posible influencia en la orientación de algunas ermitas, como por ejemplo las de La Caldereta, El Time o Vallebrón (núm 12, 30 y 36, respectivamente), todas ubicadas en barrancos que se orientan atravesando la isla y van a dar a su costa oriental. Esto incluirá un mapa topográfico detallado que permitirá poner en evidencia aspectos del terreno que en esta -nuestra primera- aproximación no llegamos a considerar. Aunque al momento no pensamos que la relevancia de la orografía se asemeje al caso ya considerado de La Gomera,[9] un estudio completo del paisaje de Fuerteventura, especialmente de la región cercana a la montaña (sagrada) Tindaya[7] o a otras chimeneas volcánicas prominentes, requiere que estos aspectos sean tenidos en cuenta.

## IV. PERSPECTIVAS DE TRABAJO FUTURO

La conquista de las canarias orientales comienza en el año 1402 comandada por los normandos Jean de Béthencourt y Gadifer de La Salle y autorizada por el rey Enrique III de Castilla. Vienen acompañados del sacerdote secular Juan de Leverrier y del monje Pedro le Bontier, líderes espirituales de la expedición.[16] Tras llegar y asentarse en Lanzarote, la expedición lleva a cabo continuas incursiones en la isla vecina. En 1404, Béthencourt y de La Salle fundan Betancuria, que se convierte en el primer asentamiento de Fuerteventura y es designada su capital. Béthencourt más tarde regresa a la Península para buscar el reconocimiento y nuevo apoyo del rey de Castilla. Pero a su regreso su aliado de

La Salle decide abandonar las islas. Aun así, en muy pocos años Fuerteventura queda controlada y la nueva sociedad se desarrolla en un periodo de convivencia entre conquistadores y aborígenes.

Desde los primeros días de la colonia, la Parroquia Matriz de la que dependía toda la isla se hallaba en Betancuria. En las décadas siguientes se fueron poblando otros centros urbanos y, lentamente, este territorio insular vio el surgimiento de haciendas y caseríos. En gran parte de estas villas, la creciente población fue acompañada por la construcción de pequeñas ermitas y templos cristianos que daban cuenta de la nueva situación religiosa y social.[14]

Como fue el caso de otros sitios del archipiélago canario, es posible que en algunos lugares las primeras ermitas se orientasen con patrones de imitación del culto prehispánico, especialmente en direcciones solsticiales.[10] Tal podría ser el caso de la ermita Nuestra Señora del Socorro en La Matilla cuyo eje se orienta con un acimut de 116,0°, prácticamente coincidente con la dirección del Sol naciente durante el solsticio de invierno boreal. Sin embargo, como ya mencionamos, esta es la única construcción histórica que se alinea de esta manera y, hasta que no surja un más firme soporte documental, ese hecho debilita el argumento. Debemos tener en cuenta, además, que el orto solar en fechas cercanas al solsticio de invierno fue una orientación muy poco usada en el mundo ibérico cristiano. Estos elementos, al momento, no nos permiten concluir nada seguro sobre posibles rastros de patrones de orientación aborigen en las iglesias que aún quedan en pie.

Nuestros resultados preliminares muestran que la gran mayoría de las ermitas e iglesias medidas (un 73% del total) se orienta con acimuts que se acomodan dentro del rango solar y, entre estas, prácticamente la totalidad (31 de las 35 construcciones canónicas) orientan sus altares hacia levante (Fig. 5). En el grupo de estas últimas se cuenta la emblemática Santa María de Betancuria, pero también varias ubicadas en otros núcleos urbanos históricos de relevancia, como por ejemplo Nuestra Señora de Antigua, y las ya mencionadas Santa Ana en Casillas del Ángel, San Antonio de Padua en Lajares y San José en el pueblo de Tesejerague.

Esta última, la ermita de San José (Tesejerague), de principios del siglo XVIII y que antes señalamos como aproximadamente equinoccial, se halla emplazada en un valle del Barranco de los Corrales y se ubica a los pies de una sierra de mediana altura. Esto hace que se vea rodeada por un horizonte montañoso, sobre todo en dirección hacia el altar (con un "h" que supera los 10°). Sabemos que un análisis completo de nuestros datos requiere tomar en cuenta la altura angular de dicho horizonte, pues un perfil elevado detrás de la iglesia cambiará la fecha en la que el Sol en el horizonte podría alinearse con su eje. Por el contrario, no esperamos que suceda algo similar con la ermita de San Francisco Javier en Las Pocetas (también equinoccial) pues su "h" medido en dirección al altar es mucho menor.

Este estudio que señalamos para San José de Tesejerague, también deberá implementarse para toda la muestra de ermitas de la isla. Un trabajo futuro, aún en progreso, nos permitirá combinar medidas locales de acimut y altura angular para estimar la declinación astronómica, coordenada que ya no dependerá de la ubicación geográfica ni de la topografía regional. El valor de esta coordenada ecuatorial, calculado para un dado templo, una vez comparado con la declinación del Sol (que fija aproximadamente un par de días en el año, o sólo uno en el caso de los solsticios), nos permitirá verificar, entre otras cosas, si esta construcción histórica está o no orientada en una dirección que coincide con la fecha de su fiesta patronal, y evaluar el peso estadístico de estos resultados. Como ya mencionamos, este análisis, al igual que una investigación más exhaustiva de las fechas de construcción de los templos en fuentes documentales, muchas aún desconocidas, está actualmente en desarrollo y esperamos poder completarlo en poco tiempo más.

En nuestro análisis consideramos también la posible influencia de la topografía como determinante del patrón de orientaciones, un elemento que ya fue verificado en estudios previos en La Gomera.[9] Sin embargo, nuestras primeras estimaciones no muestran tal correlación. Si bien puede haber casos particulares (por ejemplo, las ya mencionadas iglesias de Los Dolores o de La Merced, en La Caldereta o El Time, respectivamente), o bien el paisaje terrestre volverse relevante para aquellas ermitas en cabeceras de playa cuyos ejes son perpendiculares a la costa meridional de la isla, hemos verificado que son pocas las construcciones que se adaptan a la geografía de sus sitios de emplazamiento o que se orientan siguiendo las líneas de nivel de los barrancos. Esto nos lleva a descartar, al menos de manera preliminar, la idea de que con el correr de los años los constructores de los templos se hayan visto impelidos a modificar la tradición canónica, como conjeturamos que sí sucedió en La Gomera donde el territorio, mucho más abrupto, les jugaba en contra.[11]

Sin embargo, consideramos que un análisis completo deberá incluir otros datos, aún en estudio mediante mapas topográficos, relacionados con el perfil geográfico y la altura del horizonte circundante a los templos religiosos. Un análisis estadístico con estos elementos nos proveerá la distribución del número aproximado de iglesias por cada valor de declinación posible. Con esto podremos entonces verificar fehacientemente si la acumulación de orientaciones en el diagrama de acimuts deja su marca también en un histograma de declinaciones, que es, en fin de cuentas, aquello que nos señala la posible influencia astronómica -en particular, relacionada con el movimiento del Sol- en la orientación de las iglesias históricas.

Por último, sabemos que las islas vecinas Lanzarote y Fuerteventura, distantes en solo 10 km, están sometidas al mismo flujo de vientos, los alisios provenientes del noreste. En un trabajo previo[3] pudimos verificar que el patrón de orientaciones de Lanzarote era doble: con una alta proporción de iglesias canónicas pero, asimismo, con una componente importante y estadísticamente significativa orientada hacia el norte-noreste, que en su momento conjeturamos se debía a la intención de los constructores a soslayar las molestias causadas por la arena desplazada por el viento en las regiones cercanas al Jable.

En ese trabajo también aventuramos que quizás un patrón doble de orientaciones podría estar presente en Fuerteventura. Sin embargo, en base a nuestras mediciones y de acuerdo con los primeros resultados que hemos mostrado aquí, debemos desestimar esa idea. Nuestro análisis preliminar muestra de manera robusta que la orientación de las iglesias majoreras es insensible al flujo de vientos u otro aspecto climático que sí afectaba a su isla vecina.

## REFERENCIAS

1. A.C. González-García y J.A. Belmonte. The orientation of pre-Romanesque churches in the Iberian Peninsula. *Nexus Network Journal*, **17**: 353-377, 2015.

2. S. McCluskey. *Orientation of Christian Churches. In Ruggles, C. editor, Handbook of Archaeoastronomy and Ethnoastronomy*, New York, Springer-Verlag, 1703-1710, 2015.

3. A. Gangui, A.C. González-García, M.A. Perera Betancort y J.A. Belmonte. La orientación como una seña de identidad cultural: las iglesias históricas de Lanzarote. *Tabona. Revista de Prehistoria y Arqueología*, **20**: 105-128, 2016.

4. A. Cioranescu. *Jean de Béthencourt en Fuerteventura. En I Jornadas de Historia de Fuerteventura y Lanzarote, Tomo II*, pp. 531-546, Puerto del Rosario: Servicio de publicaciones del Cabildo de Fuerteventura, 1987.

5. C. Vogel. Sol aequinoctialis. Problèmes et technique de l'orientation dans le culte chrétien. *Revue des Sciences Religieuses*, **36**: 175-211, 1962.

6. F. Cerdeña Armas. *Noticias históricas sobre algunas ermitas de Fuerteventura. En I Jornadas de Historia de Fuerteventura y Lanzarote, Tomo I*, pp. 315-364, Puerto del Rosario: Servicio de publicaciones del Cabildo de Fuerteventura, 1987.

7. M.A. Perera Betancort, J.A. Belmonte, C. Esteban, A. Tejera Gaspar. Tindaya: un estudio arqueoastronómico de la sociedad prehispánica de Fuerteventura. *Tabona: Revista de Prehistoria y de Arqueología*, **9**: 165-196, 1996.

8. A. Gangui y J.A. Belmonte. Urban Planning in the First Unfortified Spanish Colonial Town: The orientation of the historic churches of San Cristóbal de La Laguna, *Journal of Skyscape Archaeology*, **4.1**: 6-25, 2018.

9. A. Di Paolo y A. Gangui. Estudio arqueoastronómico de las iglesias históricas de La Gomera. *Anales de la Asociación Física Argentina*, **29** (3): 62-68, 2018.

10. J.A. Belmonte, C. Esteban, A. Aparicio, A. Tejera Gaspar y O. González. Canarian Astronomy before the conquest: the pre-hispanic calendar. *Rev. Acad. Can. Ciencias*, **VI** (2-3-4): 133-156, 1994.

11. G. Díaz Padilla. La evolución parroquial de La Gomera y el patrimonio documental generado por la institución eclesiástica. *Memoria ecclesiae*, **27**: 365-376, 2005.

12. J. Concepción Rodríguez. *Fuerteventura: Obras de arquitectura religiosa emprendidas durante el siglo XVIII. En III Jornadas de Historia de Fuerteventura y Lanzarote, Tomo II*, pp. 353-383, Puerto del Rosario: Servicio de publicaciones del Cabildo de Fuerteventura, 1989.

13. B. Bonnet Reverón. Notas sobre algunos templos e imágenes sagradas de Lanzarote y Fuerteventura. *Revista de Historia Canaria*, **59**, julio-septiembre 1942.

14. A. Bethencourt Massieu. Evolución de las jurisdicciones parroquiales en Fuerteventura durante el siglo XVIII. *Revista de Historia Canaria*, **170**: 7-70, 1973.

15. M. Hernández González. *Religiosidad popular y sincretismo religioso: La Virgen de la Peña de Fuerteventura, entre lo aborigen y lo cristiano. En II Jornadas de Historia de Lanzarote y Fuerteventura, Tomo I*, pp. 195-215, Arrecife: Servicio de publicaciones del Cabildo de Lanzarote, 1990.

16. F. Caballero Mújica y M.J. Riquelme Pérez. *Santuarios marianos de Canarias*. Madrid: Encuentro, 1999.